\documentclass[conference]{IEEEtran}
\IEEEoverridecommandlockouts
\usepackage{cite}
\usepackage{amsmath,amssymb,amsfonts}
\usepackage{algorithmic}
\usepackage{textcomp}
\usepackage{xcolor}
\usepackage[pdftex]{graphicx}
\usepackage{multirow}
\def\BibTeX{{\rm B\kern-.05em{\sc i\kern-.025em b}\kern-.08em
 T\kern-.1667em\lower.7ex\hbox{E}\kern-.125emX}}
\begin{document}

\title{Lyricist-Singer Entropy Affects Lyric-Lyricist Classification Performance
}

\author{
\IEEEauthorblockN{Mitsuki Morita, Masato Kikuchi, Tadachika Ozono}
\IEEEauthorblockA{\textit{Nagoya Institute of Technology}\\
Nagoya, Aichi, Japan \\
moritsuki@ozlab.org, \{kikuchi, ozono\}@nitech.ac.jp \\}
}

\makeatletter
\def\ps@IEEEtitlepagestyle{%
  \def\@oddfoot{\mycopyrightnotice}%
  \def\@evenfoot{}%
}
\def\mycopyrightnotice{%
  \begin{minipage}{\textwidth}
  \centering \scriptsize
  Copyright~\copyright~2023 IEEE. Personal use of this material is permitted.  Permission from IEEE must be obtained for all other uses, in any current or future media, including reprinting/republishing this material for advertising or promotional purposes, creating new collective works, for resale or redistribution to servers or lists, or reuse of any copyrighted component of this work in other works.
  \end{minipage}
}
\makeatother
\maketitle

\begin{abstract}
Although lyrics represent an essential component of music, few music information processing studies have been conducted on the characteristics of lyricists. 
Because these characteristics may be valuable for musical applications, such as recommendations, they warrant further study. 
We considered a potential method that extracts features representing the characteristics of lyricists from lyrics. 
Because these features must be identified prior to extraction, we focused on lyricists with easily identifiable features.
We believe that it is desirable for singers to perform unique songs that share certain characteristics specific to the singer. 
Accordingly, we hypothesized that lyricists account for the unique characteristics of the singers they write lyrics for. 
Consequently, lyric-lyricist classification performance - or the ease of capturing the features of a lyricist from the lyrics - may depend on the variety of singers. 
In the present study, we observed a relationship between lyricist-singer entropy - or the variety of singers associated with a single lyricist - and lyric-lyricist classification performance. 
As an example, the lyricist-singer entropy is minimal when the lyricist writes lyrics for only one singer. 
We expected lyricists with small lyricist-singer entropies to be easily classifiable. 
To verify our hypothesis, we conducted the following experiments. 
First, we grouped lyricists among five groups in terms of lyricist-singer entropy and assessed the lyric-lyricist classification performance within each group. 
Subsequently, we statistically evaluated the relationship between lyricist-singer entropy and lyric-lyricist classification performance, finding a weak negative correlation that supports our hypothesis. 
Specifically, the best F1 score was obtained for the group with the lowest lyricist-singer entropy. 
Our results suggest that further analyses of the features contributing to lyric-lyricist classification performance on the lowest lyricist-singer entropy group may improve the feature extraction task for lyricists. 
\end{abstract}


\section{Introduction}

As an integral component of music, lyrics significantly influence the overall impression of a song. 
Although many studies have been conducted on the use of lyrics for song recommendation and song trend prediction tasks, lyricists have not received as much research attention as singers. 
One area of interest is authorship classification, which entails identifying the author of a text using the text itself as a source of information. 
Potential applications of authorship classification techniques include the identification of anonymous authors, as well as the prediction of attributions of criminals who issue anonymous threats. 

\begin{figure}[t]
 \centerline{\includegraphics[width=0.9\linewidth]{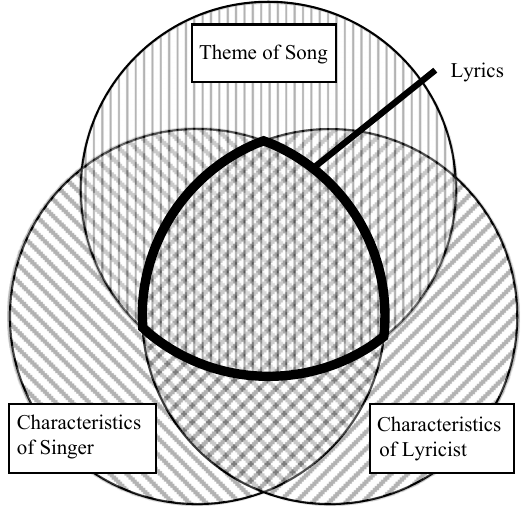}}
 \caption{Characteristics of lyrics}
 \label{image_lyric}
\end{figure}

Because lyrics are creative works authored by lyricists, we believe that the characteristics of lyricists can be leveraged in a similar manner. 
However, lyrics differ from other forms of text in that they also contain content related to a third party, namely singers. 
In other words, we believe that lyrics embody a combination of characteristics from the lyricist and singer as illustrated in Fig. \ref{image_lyric}. 

Our objective was to extract specific features that express the characteristics of lyricists from their respective lyrics. 
Nevertheless, it remains uncertain what these specific characteristics may be and to what extent they manifest in lyrics.
To address this, we conducted lyric-lyricist classification under the assumption that lyrics contain certain features. 
We implemented a classifier based on BERT \cite{BERT} that accepts lyrics as input and outputs the probabilities associated with candidate lyricists. 
Our approach is justified on the assumption that if the classifier can accurately assign multiple lyrics to the same lyricist, it can successfully capture the features associated with that lyricist. 

Initially, we considered which characteristics of lyricists who can be easily captured from lyrics.
We hypothesized that lyricists tend to account for the characteristics of singers when writing lyrics, and that lyrics based on the same characteristics would contain the same features.
Thus, we hypothesized that it would be more difficult to classify lyrics from multiple singers to one lyricist than to classify lyrics from one singer to one lyricist.
To verify this hypothesis, we quantified the variety of singers associated with each lyricist as a measure of lyricist-singer entropy and the ease of capturing features as a measure of lyric-lyricist classification performance. 
We then evaluated a relationship between lyricist-singer entropy and lyric-lyricist classification performance by grouping lyricists based on their lyricist-singer entropy and calculating lyric-lyricist classification performance for each group. 
As a result, we found a negative correlation between lyricist-singer entropy and lyric-lyricist classification performance, which supports our hypothesis. 

\section{Related Work}
\subsection{Authorship Classification}
Authorship classification \cite{survay_AA}, the classification of authors from their texts, has been an objective of numerous studies. 
One approach to this task is to focus on unique textual expressions, also known as idiolect.
Mael et al. \cite{AC_BertAA} proposed a BERT-based authorship ensemble classification model that combines outputs from logistic regression models utilizing stylometric features based on econometric literature concepts, as well as hybrid features calculated from N-grams. 
Dhar \cite{AC_CNN} developed an authorship classification system using a convolutional neural network (CNN), which processes natural language sentences by converting them into sentence vectors prior to feature extraction and classification. 
Sari et al. \cite{AC_feature} performed authorship classification through logistic regression using simple sentence features, including style features that represent the use of function words, numbers, and punctuation, content features derived from word N-grams in sentences excluding function words, and hybrid features combining both feature types. 
Fedotova et al. \cite{AA_fanfiction} conducted authorship classification in the domain of anonymous fanfiction texts. 
Fanfiction, which refers to content written by fans of existing works, is easy to treat as a theme because it shares characteristics with the works it is derived from. 
In the study, features were extracted using fastText and classified using SVM. 
Corbara et al. \cite{AA_prosodic} proposed the prosodic clause, or the number of sounds in a word, as a feature in prose that does not specify a format. 
Prosodic clauses are hypothesized to represent an author's features irrespective of textual meaning. 

Because our objective was to extract the features of lyricists (authors) from lyrics (texts), we considered approaching this task as idiolect-based authorship classification.

\subsection{Lyrics Analysis}
In this study, we performed lyric-lyricist classification as a method to analyze lyrics. 
Because the analysis of lyrics may serve various purposes, such as song recommendation and trend prediction, many analytical methods have been proposed. 
Velankar et al. \cite{LA_mood} analyzed Hindi lyrics written in a script called Devanagari by defining five moods. 
Corbara et al. \cite{LA_embedding} classified genres and artists by incorporating information from the Billboard magazine, including the durations and rankings of songs published in the magazine. 
Haraguchi et al.\cite{LA_LyricVideo} analyzed lyrics videos, which are often used as promotional content, focusing on three aspects: text movement, font style, and music style.
Yılmaz et al. \cite{LA_AA} analyzed lyrics-specific expressions, such as repetitive expressions and vocabulary, by employing authorship classification, occlusion analysis, and genre classification using GloVe word embeddings, pre-trained subword-level embeddings, and phoneme encodings.

\section{Lyricist-Singer Entropy}
We hypothesized that lyricists consider the characteristics of the singers they write lyrics for, which may impact lyric-lyricist classification performance.
To validate this hypothesis, we quantified the variety of singers associated with each lyricist as lyricist-singer entropy. 
The following subsections present notations used throughout the text, describe the collection of relevant data and discuss the methods employed to calculate lyricist-singer entropy.

\subsection{Notation}
Letting $X$ be a set of songs, a song $x$ is defined as $x=\langle i,~j,~s\rangle$, where $i \in I$, $j \in J$, and $s$ denote the lyricist, singer, and lyrics of $x$, respectively, with $s$ representing textual data.

We defined the lyricist-singer entropy as a characteristic associated with each lyricist, and grouped lyricists among five groups according to this characteristic. 
The lyricist-singer entropy of lyricist $i$ is denoted as $H_i$. 
Because we employed two different grouping methods, we use $A$ and $B$ to denote the two respective sets of lyricist groups, where $A_k$ denotes the $k$-th group of $A$.
We use the notation $\hat{A_k}$ to denote the extraction of a subset from $A_k$. 

Each song in our dataset is represented by lyrics and a corresponding one-hot vector indicating the actual lyricist. 
We denote the overall dataset as $D$, the set of candidate lyricists in $D$ as $D_I$, and the set of songs in $D$ as $D_X$. 
The number of elements in any set $X$ is denoted as $|X|$.

\subsection{Collecting Song Information}
We collected song information from the lyrics search service Uta-Net\footnote{https://www.uta-net.com}, which assigns unique IDs to each song, lyricist, and singer, and provides individual pages for each song where this information can be found. 
We state that lyricist $i$ wrote lyrics to singer $j$ when there exists a song $x=\langle i, j, s\rangle$. 
For example, songs $x_1 = \langle 10, 20, s_1\rangle$ and $x_2 = \langle 10, 30, s_2\rangle$ were written by lyricist 10 and performed by singers 20 and 30, respectively. 

During the information collection period from April 25, 2022, to May 1, 2022, there were approximately 300,000 song pages on Uta-net. 
At first, we randomly selected 30,000-IDs without duplication. 
Subsequently, we excluded songs written by lyricists with fewer than ten songs, as the number of songs was considered insufficient to include in a lyric-lyricist classification dataset. 
All remaining songs were included in dataset $X$ with the following size. 
$|X| = 10444$, $|I| = 499$, and $|J| = 3300$. 

To address concerns regarding the inconsistent representation of lyricist names on Uta-Net, we conducted a verification using the Levenshtein distance, which represents the number of insertion, deletion, and conversion operations required to convert one string to another. 
After listing all pairs of lyricists with a Levenshtein distance of 1, we conducted a manual review to confirm that none of these lyricists represented the same person with a slight variation in names. 

\subsection{Calculation of Lyricist-Singer Entropy}

\begin{figure}[t]
 \centerline{\includegraphics[width=\linewidth]{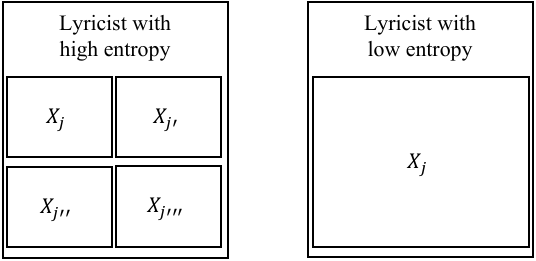}}
 \caption{Concept of lyricist-singer entropy}
 \label{other_singer}
\end{figure}

Fig. \ref{other_singer} illustrates the concept of lyricist-singer entropy. 
For example, if lyricist $i$ has written lyrics to four different singers $j$, $j'$, $j''$, and $j'''$, they would be considered to have a high lyricist-singer entropy, associated with four sets of lyrical characteristics. 
Conversely, if lyricist $i$ has exclusively written lyrics to a specific singer $j$, they would have a minimal lyricist-singer entropy, and any songs written by $i$ would reflect the unique characteristics of both the singer and lyricist. 
We note that neither scenario depicted in the figure guarantees that the set of songs for a singer is a subset of the set of songs for a lyricist; i.e., the same singer may be associated with different lyricists. 

Lyricist-singer entropy represents the probabilistic entropy that a song written by a certain lyricist is sung by each singer.
We calculate the lyricist-singer entropy $H_i$ for a lyricist $i$ using the following formula: 

\begin{eqnarray*}
 H_i = - \sum_{j \in J}\frac{|X_i \cap X_j|}{|X_i|} \log \frac{|X_i \cap X_j|}{|X_i|},
\end{eqnarray*}
where $|X_i \cap X_j|$ represents the number of songs included in both $X_i$ (set of songs written by lyricist $i$) and $X_j$ (set of songs performed by singer $j$), and $|X_i|$ represents total number of songs written by lyricist $i$. 

\begin{figure}[t]
 \centerline{\includegraphics[width=\linewidth]{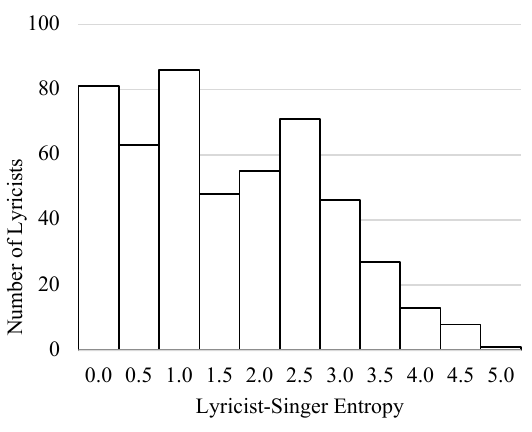}}
 \caption{Histogram of lyricist-singer entropy contained in lyricists' set $I$}
 \label{entropy_distribution}
\end{figure}

Fig. \ref{entropy_distribution} presents the distribution of lyricist-singer entropy for lyricists in the set $I$. 
The vertical axis represents the number of lyricists, and the horizontal axis represents the lyricist-singer entropy value. 

\section{Lyric-Lyricist Classification}
Lyric-lyricist classification is a task wherein a classifier accepts lyrics as input, calculates the probabilities of each candidate lyricist with respect to said input, and predicts the lyricist who wrote the input lyrics as output. 
First, we grouped lyricists according to lyricist-singer entropy based on the assumption that lyricists within a certain range of entropy exhibit similar characteristics. 
Next, we constructed datasets for lyric-lyricist classification. 
Finally, we calculated the lyric-lyricist classification performance for each lyricist, as well as the average lyric-lyricist classification performance within each group. 
The following subsections present details pertaining to the grouping methods, lyric-lyricist classifiers, dataset construction, and experimental results. 

\subsection{Grouping Lyricists}
To avoid potential biases, we employed two different methods to group lyricists, with the respective sets of groups denoted as $A$ and $B$. 
In both cases, we assigned lyricists with a lyricist-singer entropy of 0 to individual groups $A_0$ and $B_0$, considering them as unique cases. 
All other lyricists were assigned to groups $A_{1-4}$ or $B_{1-4}$ in ascending order of lyricist-singer entropy.
Because we employed the F1 score as a measure of lyric-lyricist classification performance, it was desirable to construct more groups.
However, as shown in Fig. \ref{entropy_distribution}, the number of lyricists substantially decreased for lyricist-singer entropy measures exceeding 3.
Hence, we concluded that five groups would be appropriate for constructing unbiased testing datasets and performing reliable experiments.
To construct group set $A$, we ensured that each group had an equal number of lyricists. In contrast, to construct group set $B$, we minimized the lyricist-singer entropy variance within each group.

The first method assigns an equal number of lyricists to $A_{1-4}$ in decreasing order of lyricist-singer entropy.
For instance, if there are $n$ lyricists with non-zero lyricist-singer entropy, lyricists are assigned to groups $A_k$ with lyricist-singer entropies in the range of $n(k - 1) / 4 + 1$ to $nk / 4$ to $A_k, k=1,2,3,4$, with fractional values being truncated.
For the second method, we employed $k$-means clustering by lyricist-singer entropy with $A_k$ as initial states to minimize the lyricist-singer entropy variance within each group $B_k, k = 1, 2, 3, 4$. 

\begin{table}[t]
 \caption{Statistics for each lyricist group}
 \begin{center}
 \begin{tabular}{c|c|cc|cc}
 \hline
 \multirow{2}{*}{Group}&Number of&
 \multicolumn{2}{c|}{Number of Songs}&
 \multicolumn{2}{c}{Lyricist-Singer Entropy}\\
 & Lyricists & Average & Total & Average & Range \\
 \hline
 $A_0$ & 81 & 13.370 & 1,083 & 0.000 & 0.000 - 0.000\\
 $A_1$ & 104 & 15.865 & 1,650 & 0.428 & 0.146 - 0.670\\
 $A_2$ & 105 & 18.638 & 1,957 & 1.101 & 0.679 - 1.666\\
 $A_3$ & 104 & 15.692 & 1,632 & 2.065 & 1.666 - 2.458\\
 $A_4$ & 105 & 39.257 & 4,122 & 3.108 & 2.458 - 4.583\\\hline
 $B_0$ & 81 & 13.370 & 1,083 & 0.000 & 0.000 - 0.000\\
 $B_1$ & 177 & 16.633 & 2,944 & 0.691 & 0.146 - 1.232\\
 $B_2$ & 115 & 17.035 & 1,959 & 1.840 & 1.245 - 2.272\\
 $B_3$ & 95 & 21.463 & 2,039 & 2.725 & 2.303 - 3.233\\
 $B_4$ & 31 & 78.032 & 2,419 & 3.778 & 3.284 - 4.583\\
 \hline
 \end{tabular}
 \label{group_statistics}
 \end{center}
\end{table}
TABLE \ref{group_statistics} provides statistics of groups $A_k$ and $B_k$. 
The number of lyricists represents a total number of lyricists in each group. 
For instance, $A_1$ contains 104 lyricists. 
The number of songs represents the average or total number of songs of all lyricists in each group. 
The average is calculated as $\sum_{i \in A_k}|X_i| / |A_k|$, and the total is calculated as $\sum_{i \in A_k}|X_i|$, where $X_i$ represents set of songs written by lyricist $i$. 
The lyricist-singer entropy represents lyricist-singer entropy for all lyricists in each group. 
The average is calculated as $\sum_{i \in A_k} H_i / |A_k|$, where $H_i$ represents the lyricist-singer entropy of lyricist $i$. 
The range is represented by the minimum value $\min\limits_{i \in A_k} H_i$ and maximum value $\max\limits_{i \in A_k} H_i$. 
The same statistics are provided for the groups $B_k$. 

\subsection{Classifier}
Our classifier was developed based on a BERT model pre-trained on Japanese texts\footnote{https://github.com/cl-tohoku/bert-japanese.git}. 
We fine-tuned the linear and final layers of the models. 
The input and output of the fine-tuned model are tokenized lyrics and a lyricist vector of length 10, as ten lyricists are included in the candidate set. 
The $k$-th element of the lyricist vector denotes the probability that the corresponding lyricist wrote the input text.
Our model accepts the first 512 tokens as input due to the limitation of BERT. 
We decided there would be little impact on the result because almost all lyrics in our dataset are within a length of 512 tokens or less. 

For each dataset, a separate model was constructed, trained, and deployed. 
All training, validation, and testing procedures were performed using data represented as pairs of lyrics and one-hot vectors indicating the position of the true lyricist among the candidates. 

During model training, lyrics were input into the model to obtain outputs, which were then passed to the loss function along with the one-hot vectors for model updates. 
After the entire training dataset was processed a similar procedure was performed with the validation dataset to obtain and record the loss function.
An epoch was defined as one complete pass of processing both the training and validation datasets. 
An early stopping mechanism was implemented to terminate the training process and save the model when the loss function from the validation dataset surpassed the previous epoch's loss function three times.

\subsection{Evaluation Metrics}
First, we calculated the precision, recall, and F1 score as evaluation metrics for the lyric-lyricist classification performance of lyricist $i$.
Next, we calculate the averages of each metric within each group.
For instance, the values $Precision_0$, $Recall_0$, and $F_0$ denote the average precision, recall, and F1 score, respectively, of the lyricists in $A_0$.
Higher values within these metrics indicate higher lyric-lyricist classification performance, which in turn suggests the presence of common lyrical features among songs of the lyricist.

\subsection{Dataset}
We constructed datasets of 100 songs and corresponding candidate lyricist vectors, with ten songs per lyricist. 
For each lyricist within a dataset, we allocated six songs for training, two songs for evaluation, and two songs for testing.
The construction process of a dataset encompassed two stages: selecting the lyricists and selecting the songs. 
The selected lyricists are denoted as $D_I$, and some lyricists in the lyricist set $A_k$, $B_k$ are denoted as $\hat{A_k}$, $\hat{B_k}$, respectively. 
To examine the effect of sampling on classification results, we employed two sampling methods for dataset construction: homogenous sampling, and heterogenous sampling.

In homogenous sampling, ten lyricists were randomly selected from the same group. 
For example, if we select ten lyricists from $A_0$, we have $D_I = \hat{A_0}$ and $|\hat{A_0}|=10$.
During the experiment, this process was repeated ten times for each group.

In heterogenous sampling, we randomly selected two lyricists from each group. 
If we construct a dataset from group $A$, the candidate lyricists in dataset $D_I$ are given by $D_I = \cup_{k = 0}^{4}\hat{A_k}$, where $|\hat{A_k}| = 2, k = 0, 1, 2, 3, 4$.
During the experiment, this process was repeated 50 times for each group. 

\subsection{Result}
\begin{figure}[t]
 \centerline{\includegraphics[width=\linewidth]{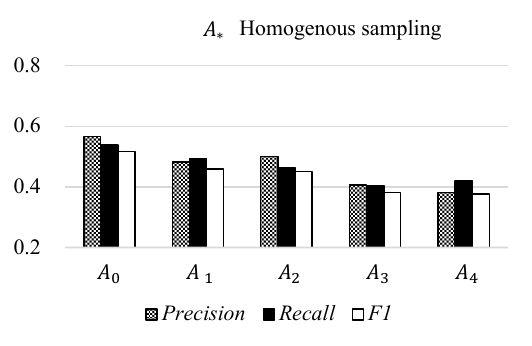}}
 \caption{Classification performance on homogenous sampling $A_*$}
 \vspace{3mm}
 \label{order_same}
\end{figure}

\begin{figure}[t]
 \centerline{\includegraphics[width=\linewidth]{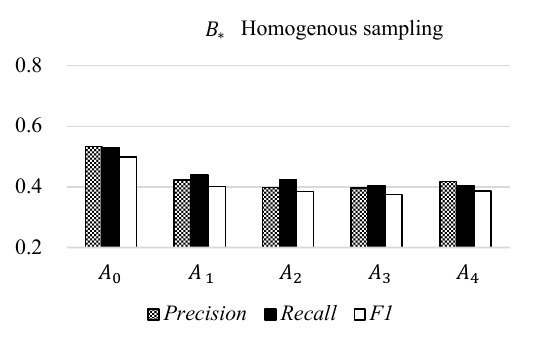}}
 \caption{Classification performance on homogenous sampling $B_*$}
 \label{k_means_same}
\end{figure}

\begin{figure}[t]
 \centerline{\includegraphics[width=\linewidth]{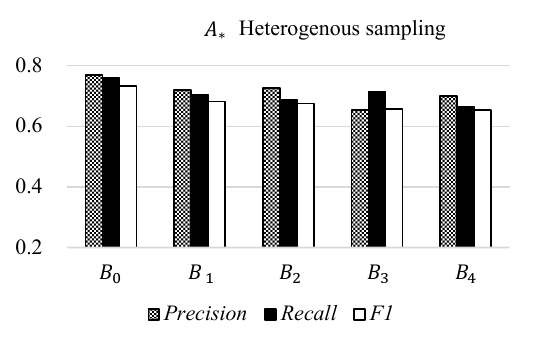}}
 \caption{Classification performance on heterogenous sampling $A_*$}
 \vspace{3mm}
 \label{order_separate}
\end{figure}

\begin{figure}[t]
 \centerline{\includegraphics[width=\linewidth]{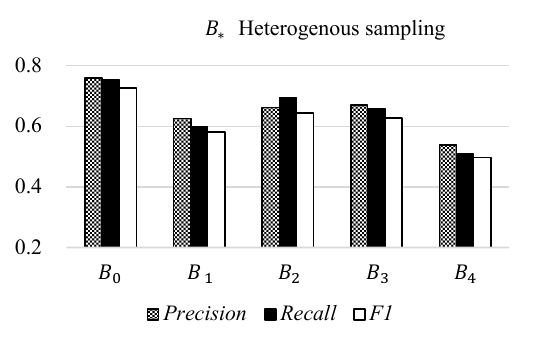}}
 \caption{Classification performance on heterogenous sampling $B_*$}
 \label{k_means_separate}
\end{figure}

Figs. \ref{order_same}-\ref{k_means_separate} and Tables \ref{order_same_value}-\ref{k_means_separate_value} present lyric-lyricist classification performance results for subsets of $A$ and $B$ constructed under different sampling strategies.

We expected lyric-lyricist classification performance to exhibit a negative correlation with lyricist-singer entropy.
The presented results support this hypothesis, although the variability in performance between the groups is small.

\section{Discussion}

We analyzed the relationship between lyricist-singer entropy and lyric-lyricist classification performance, where the former is a measure of singer diversity for an individual lyricist, and the latter is calculated through the lyric-lyricist classification task defined in this study. 
We found a negative correlation between lyricist-singer entropy and lyric-lyricist classification performance. 

We used two methods of grouping and two methods of dataset construction. 
The results for all cases demonstrate that groups with higher lyricist-singer entropy exhibited lower lyric-lyricist classification performance. 
In particular, $A_0$ and $B_0$, consisting of lyricists with a lyricist-singer entropy of 0, were associated with significantly better lyric-lyricist classification performance than the other groups in both cases. 
Lyricists with high lyricist-singer entropy write lyrics for a variety of singers, often without singing the lyrics themselves. 
The negative correlation between lyricist-singer entropy and lyric-lyricist classification performance suggests that lyricists who write lyrics for a variety of singers may prioritize content pertaining to the singer, rather than their own characteristics. 
Conversely, lyricists who write lyrics for one particular singer may emphasize their own characteristics in their lyrics. 

We conducted experiments using two cases: a case wherein songs were written by lyricists with similar lyricist-singer entropy measures (homogenous sampling), and a case wherein songs were written by lyricists with significantly different lyricist-singer entropy measures (heterogenous sampling). 
Because we hypothesized that it is easier to capture lyricist features given a lower lyricist-singer entropy, we expected higher performance for homogenous sampling with lower lyricist-singer entropy groups. 
Conversely, we expected higher lyric-lyricist classification performance for heterogenous sampling with higher lyricist-singer entropy groups. 
However, we found that heterogenous sampling yielded higher lyric-lyricist classification performance than heterogenous sampling irrespective of in-group lyricist-singer entropy, which we could not explain. 

We used two methods to lyricists into sets $A$ and $B$, based on the number of lyricists and lyricist-singer entropy, respectively, to reduce bias in the number of people or the characteristics of the lyricists in a group according to the method of grouping. 
In $B$, the number of lyricists in $B_1$ was approximately twice that in the other groups.
Furthermore, the lyric-lyricist classification performance of $B_1$ was lower than that of $B_2$ and $B_3$ under heterogenous sampling, which deviates from the overall tendency.
This may indicate that the lyricists were not appropriately grouped. 

Ultimately, we found that lyricists who write to a smaller number of singers exhibit easily identifiable features.
However, our experimental setup was limited in certain aspects. 
Future studies may involve conditioning variability between the groups, as well as a larger dataset to ensure a more detailed grouping.

\section{Conclusion}
\begin{table}[t]
 \caption{Lyric-lyricist classification performance \\on homogenous sampling $A_*$}
 \begin{center}
 \begin{tabular}{c|ccc}
 \hline 
 & $Precision$ & $Recall$ & $F1$ \\\hline
 $A_0$ & 0.566 & 0.540 & 0.517 \\
 $A_1$ & 0.482 & 0.495 & 0.460 \\
 $A_2$ & 0.501 & 0.465 & 0.451 \\
 $A_3$ & 0.408 & 0.405 & 0.383 \\
 $A_4$ & 0.382 & 0.420 & 0.378 \\
 \hline 
 \end{tabular}
 \label{order_same_value}
 \end{center}
\end{table}

\begin{table}[t]
 \caption{Lyric-lyricist classification performance \\on homogenous sampling $B_*$}
 \begin{center}
 \begin{tabular}{c|ccc}
 \hline 
 & $Precision$ & $Recall$ & $F1$ \\\hline
 $B_0$ & 0.534 & 0.530 & 0.499 \\
 $B_1$ & 0.422 & 0.440 & 0.401 \\
 $B_2$ & 0.399 & 0.425 & 0.385 \\
 $B_3$ & 0.397 & 0.405 & 0.376 \\
 $B_4$ & 0.419 & 0.405 & 0.387 \\
 \hline 
 \end{tabular}
 \label{k_means_same_value}
 \end{center}
\end{table}

\begin{table}[t]
 \caption{Lyric-lyricist classification performance \\on heterogenous sampling $A_*$}
 \begin{center}
 \begin{tabular}{c|ccc}
 \hline 
 & $Precision$ & $Recall$ & $F1$ \\\hline
 $A_0$ & 0.769 & 0.760 & 0.733 \\
 $A_1$ & 0.722 & 0.705 & 0.682 \\
 $A_2$ & 0.727 & 0.690 & 0.677 \\
 $A_3$ & 0.656 & 0.715 & 0.658 \\
 $A_4$ & 0.700 & 0.665 & 0.653 \\
 \hline 
 \end{tabular}
 \label{order_separate_value}
 \end{center}
\end{table}

\begin{table}[t]
 \caption{Lyric-lyricist classification performance \\on heterogenous sampling $B_*$}
 \begin{center}
 \begin{tabular}{c|ccc}
 \hline 
 & $Precision$ & $Recall$ & $F1$ \\\hline
 $B_0$ & 0.761 & 0.755 & 0.728 \\
 $B_1$ & 0.626 & 0.600 & 0.582 \\
 $B_2$ & 0.662 & 0.695 & 0.645 \\
 $B_3$ & 0.671 & 0.660 & 0.628 \\
 $B_4$ & 0.540 & 0.510 & 0.497 \\
 \hline 
 \end{tabular}
 \label{k_means_separate_value}
 \end{center}
\end{table}

This study was conducted to examine how lyricist-singer entropy affects lyric-lyricist classification performance. 
We found that lyricists with a lower lyricist-singer entropy tend to be easier to classify, and lyricists with a lyricist-singer entropy of 0 are significantly easier to classify.
We conducted lyric-lyricist classification experiments to evaluate the relationship between lyricist-singer entropy and lyric-lyricist classification performance. 
Our hypothesis states that classifying lyrics with different singers to the same lyricist is more challenging than classifying the lyrics of one singer to the same lyricist. 
The experimental results demonstrate weak support for our hypothesis. 
Further analysis of lyrics written by lyricists with a lyricist-singer entropy of zero may be promising in interpreting the features contributing to lyric-lyricist classification performance.
\section*{Acknowledgment}
This work was supported in part by JSPS KAKENHI Grant Numbers JP19K12266, JP22K18006.
\newpage

\bibliographystyle{unsrt}
\bibliography{morita_2023}

\end{document}